\documentclass[10pt,preprint]{aastex}
\usepackage{graphicx}

\received{7/29/2002}
\revised{11/25/2002}
\accepted{12/06/2002}

\def\twodust{%
{\bf 2-D}{\sc ust}}
\def\iso{{\sl ISO}}
\def\iras{{\sl IRAS\/ }}
\def\rstar{R_{*}}
\def\rsun{R_{\odot}}
\def\teff{T_{\rm eff}}
\def\rmin{r_{\rm min}}
\def\rmax{r_{\rm max}}
\def\nrad{n_{r}}
\def\nlat{n_{\Theta}}
\def\ntheta{n_{\theta}}
\def\nphi{n_{\phi}}
\def\tdust{T_{\rm dust}}

\def\lstar{L_{*}}
\def\thincl{\theta_{\rm incl}}
\def\um{\mu {\rm m}}
\def\90deg{90^{\circ}}
\def\45deg{45^{\circ}}
\def\0deg{0^{\circ}}
\makeatletter
\def\ale{\mathrel{\mathpalette\gl@align<}}
\def\age{\mathrel{\mathpalette\gl@align>}}
\def\gl@align#1#2{\lower.6ex\vbox{\baselineskip\z@skip\lineskip\z@
\ialign{$\m@th#1\hfil##\hfil$\crcr#2\crcr\sim\crcr}}}
\makeatletter

\shorttitle{Dust Radiative Transfer for an Axisymmetric System} 
\shortauthors{Ueta \& Meixner}

\begin{document}
 
\title{{\protect\twodust}: A Dust Radiative Transfer Code for an 
Axisymmetric System}

\author{Toshiya Ueta\altaffilmark{1} and Margaret Meixner\altaffilmark{2}}

\affil{Department of Astronomy, MC-221, 
University of Illinois at Urbana-Champaign, 
Urbana, IL  61801}

\altaffiltext{1}{Current Address:
Royal Observatory of Belgium,
Avenue Circulaire, 3, 
B-1180, Bruxelles, Belgium;
ueta@oma.be}

\altaffiltext{2}{Current Address:
Space Telescope Science Institute, 
3700 San Martin Drive, Baltimore, MD 21218, USA;
meixner@stsci.edu}

\begin{abstract}
We have developed a general purpose dust radiative transfer 
code for an axisymmetric system, {\twodust}, motivated by 
the recent increasing availability of high-resolution images 
of circumstellar dust shells at various wavelengths.
This code solves the equation of radiative transfer following
the principle of long characteristic in a 2-D polar grid 
while considering a 3-D radiation field at each grid point.
A solution is sought through an iterative scheme in which 
self-consistency of the solution is achieved by requiring a 
global luminosity constancy throughout the shell.
The dust opacities are calculated through Mie theory from 
the given size distribution and optical properties of the dust 
grains.
The main focus of the code is to obtain insights on 
(1) the {\sl global} energetics of dust grains in the shell
(2) the 2-D projected morphologies that are strongly dependent 
on the mixed effects of the axisymmetric dust distribution 
and inclination angle of the shell.
Here, test models are presented with discussion of the results.
The code can be supplied with a user-defined density distribution 
function, and thus, is applicable to a variety of dusty 
astronomical objects possessing the axisymmetric geometry.
\end{abstract}


\section{Introduction}

Astronomical systems are often surrounded by a shroud
of dust.
Evolved stars are the most typical of such, since they 
are responsible for more than 80\% of the material 
annually injected into the interstellar space through 
dusty mass loss \citep{sedlmayr94}.
The ejected matter forms a dust-rich shell around these 
stars, which can be very bright in the mid-infrared
(mid-IR; $\sim 10 - 20 \mu$m) due to thermal emission 
from warm (a few 100 K) dust grains.
Therefore, the dust distribution in these circumstellar 
shells can be directly probed in the mid-IR.

Recent mid-IR observations at dust continuum 
have revealed toroidal density distribution in the 
circumstellar shells of evolved stars 
(e.g., \citealt{skinner94,dayal98,meixner99}).
Such axisymmetric dust distributions have been seen 
not only in the circumstellar shells of evolved 
stars but also in the shells of massive, young 
stars (e.g., \citealt{ueta01b,smith02}).
Most recently, the use of large aperture telescopes 
with mid-IR capabilities has pushed the 
diffraction-limited mid-IR imaging to sub-arcsecond 
resolution, and the intrinsically compact structure 
of the circumstellar dust shells has been revealed
(e.g., \citealt{jura01,ueta01a}).

However, it is not easy to interpret the mid-IR
images since the mid-IR morphologies of these dust 
shells are highly influenced by self-extinction
introduced by the geometry and inclination of 
the axisymmetric shells.
Therefore, we need to construct numerical models 
in more than 1-D to properly interpret
high-resolution mid-IR images of the 
circumstellar shells and fully understand the 
intertwined relationship between the morphologies 
and the dust distribution.

Furthermore, recent {\iso} observations have 
greatly enhanced our knowledge of the circumstellar 
dust mineralogy 
(e.g., \citealt{waters96,kemper02,molster02}).
Together with the increasing availability of 
laboratory-measured optical constants for astronomical 
dust analogs 
(e.g., \citealt{jager94} and subsequent series of papers;
\citealt{jager98}; \citealt{speck98} for a recent 
compilation), a more elaborate treatment of dust grains
is necessary to properly model the energy budget
within the circumstellar dust shells, especially when 
dust grains are the primary means for energy transport
(e.g., \citealt{ueta01a,ueta01b,meixner02}). 

In this context, we have developed a general purpose
radiative transfer code, {\twodust}, for an axisymmetric 
dust system.
Below, we will introduce the code, mainly focusing on 
the treatment of dust grains (\S \ref{method}; also 
see appendices), discuss the 
results of test models (\S \ref{run}), and give a 
summary (\S \ref{2dustconcl}).

\section{The {\protect\twodust} code\label{method}}

\subsection{Brief Overview}

The {\twodust} code solves the equation of radiative 
transfer and derive the radiation and temperature 
field within a 2-D polar grid, while considering 
a fully 3-D radiation field.
The code is based on the iterative scheme elucidated 
by \citet{collison91} using the principle of long 
characteristic and is written in Fortran 90 to 
allow dynamic memory allocation for parameter arrays.
The computational algorithms and assumptions are
outlined in Appendix \ref{detailmethod}.
We have chosen the long characteristic method
over other 2-D radiative transfer methods such
as the monte Carlo method (e.g., \citealt{ldb83}),
the short characteristic method (e.g., \citealt{ka88}),
and the moment method (e.g., \citealt{spagna91}),
because of the method's simplicity and 
straightforwardness in implementation
(see \citet{dullemond00} for a discussion on the
pros and cons for each method).
This method has not been widely used because of 
its tendency to be computationally expensive. 
However, this problem can be alleviated by the
parallelization of the code exploiting the
heavily looped structure of the algorithm.

Our unique approach is to recognize the inner 
radius of the circumstellar shell as an observable 
that can be measured from high-resolution mid-IR 
images (e.g., \citealt{ueta01a,ueta01b,meixner02}).
Once the inner shell radius is observationally 
determined, the dust temperature at the inner 
radius can be specified almost immediately.
Then, the subsequent derivation of the temperature 
and radiation field within the shell is relatively 
straightforward.

The inner shell radius may alternatively be fixed 
by assuming the dust temperature at the inner radius 
to be equal to the dust condensation temperature
(e.g., \citealt{efstathiou90,men97}), for example.
However, this condition is not necessarily true
when the dust shell is physically detached from 
the central source, since in such a case the inner 
edge of the shell does not correspond to the dust 
condensation radius.
Our approach is general and does not require any 
assumption: the inner shell radius may correspond to 
the dust condensation radius as in the dust-forming
circumstellar wind shells, a precipitous density drop 
due to cessation of mass loss as in the detached shells, 
or the swept-up shell boundary caused by a sudden mass 
ejection.

Then, we iterate on the model parameters 
by using the spectral energy distribution (SED) 
{\sl and} the mid-IR images as constraints.
The measured inner shell radius is a very strong
constraint on the energetics within the dust shell,
and helps to investigate the dust mineralogy 
(composition and size distribution) with sufficient 
details.
Moreover, the mid-IR images themselves do constrain
the axisymmetric dust distribution of the model, and
would aid to disentangle the combined effects of the 
optical depth and the inclination angle of the shell
to the projected shell morphologies.

\subsection{Treatment of Dust Grains\label{treat}}

One of the most crucial parts of the radiative 
transfer in a dusty medium is proper considerations 
of the dust cross sections.
Our aim with {\twodust} is to model the dust continuum 
emission from the axisymmetric shell and to gain 
insights on the {\sl global} energetics of the dust
shell for a wide wavelength range between ultraviolet 
and far-IR.
Therefore, we compute cross sections for a fiducial 
dust species that exhibits the ``averaged'' optical 
properties of all the dust species present in 
the shell instead of following each dust component 
to reproduce each of the specific narrow dust features.

Three assumptions that come into our dust consideration 
are that
(1) all of the dust species are well-mixed (i.e., homogeneous), 
(2) dust grains are well-equilibrated with the radiation 
field (i.e., single dust temperature for all dust species), and
(3) dust grains are spherical particles.
The latter two assumptions may not reflect the reality
very well especially since small dust grains are known 
to be transiently heated (e.g., \citealt{skm92})
and it is more realistic to consider a distribution 
of ellipsoidal shapes (e.g., \citealt{bohren83}).
These issues are out of the scope for the present
study and will be addressed in the future upgrade 
of the code.
Using the laboratory-measured refractive index,
we calculate ``$Q$'' efficiency factors for the 
extinction, scattering, and absorption cross 
sections of the dust particles through Mie theory
\citep{hulst57,bohren83}.
Since the $Q$ factors are size and frequency 
dependent, we integrate over the size space
at each wavelength (frequency) grid.
We adopt two dust size distributions derived from
the study of the interstellar medium 
(\citealt{mathis77}; hereafter the ``MRN'' 
distribution) and the study of the interstellar 
medium and the circumstellar shells
(\citealt{kim94,jura94}; hereafter the ``KMH'' 
distribution).

Only isotropic scattering is typically considered
in most dust radiative transfer codes.
However, this simplification may not be appropriate 
when there are large dust grains that are known to 
forward scatter (e.g., \citealt{bohren83}). 
The presence of large grains has been suggested to 
explain, for example, the observed circumstellar 
polarization at the $K$ band around a carbon-rich 
asymptotic giant branch (AGB) star, IRC +10216 
\citep{jura94} and the observed 
millimeter-wave excess in the SED of the circumstellar 
disk around a T Tauri star, TW Hydra \citep{wein02}.
Therefore, we have incorporated anisotropic scattering 
by generalizing the source function with the modified 
Henyey-Greenstein phase function \citep{cs92}.
This phase function has been selected because 
(1) the function has a simple two-parameter analytic form, and
(2) the function is physically reasonable.
We assume azimuthal symmetry of scattering with 
respect to the angle of incident because dust grains
are unlikely to scatter incident radiation into a
specific azimuthal direction preferentially over other 
azimuthal directions under the assumption of randomly 
oriented dust grains.

\section{Results of the \protect\twodust\ Code\label{run}}

The {\twodust} code requires a large number of input 
parameters to be supplied upon execution.
These parameters can generally be divided into three
categories related to
(1) the computational grid, 
(2) the physical nature of the dust shell system, and
(3) the dust grain properties.
Table 1 summarizes the input parameters.
There are also a large number of output values
generated from the code that are summarized in Table 2.
The most important is a list of the specific intensities 
($J_{\nu}\/$ when isotropic or $I_{\nu}$ when anisotropic,
where $\nu$ is frequency) 
and dust temperatures at each grid point, from which
the SED of the model and two dimensional
projected surface brightness and optical depth maps
can be generated for a given inclination angle at a given
wavelength (frequency).

\subsection{Spherically Symmetric Models\label{spherical}}

We tested the {\twodust} code by constructing a number of
spherically symmetric models and comparing the results with 
results of a 1-D radiative transfer code, {\sf DUSTY} 
\citep{dusty}.
As a test case, we considered a circumstellar shell of the 
$r^{-2}$ density profile, surrounding an F1 post-AGB 
central star (see Table \ref{twodust_tab1} for parameters).
The shell is assumed to be composed of amorphous silicate 
grains (Olivine; olmg50 in \citealt{dorschner95}).
Then, we varied the optical depth of the shell and the dust 
size distributions under the isotropic and anisotropic 
scattering assumptions

The spherical shell models of the two codes agreed quite 
well.
The surface brightness maps showed the radial profile of 
$r^{-2}$ as expected from dust-scattering of star light 
for the $r^{-2}$ density distribution.
Especially good agreement was seen in the radial dependence 
of the temperature structure.
The {\twodust} results, however, yielded slightly higher 
temperature than the {\sf DUSTY} results (at most about 
10\% difference), which was due to the curvature effect 
at the inner edge of the shell:
3-D accounting of available radiation in {\twodust} has 
resulted in a slightly larger flux density and dust 
temperature.

The main difference between the MRN and KMH size 
distributions is the existence of the maximum grain 
size in the MRN distribution.
Hence, the MRN distribution tends to have larger 
weights on the small grain population, making the MRN 
grains more absorptive than the KMH grains.
This generally yields lower optical flux in the MRN 
models than the KMH models.
The surface brightness maps also showed this trend:
surface brightness at the optical and near-IR was 
generally smaller in the MRN models than the KMH models.

The models with the anisotropic scattering grains
showed the forward scattering nature of large dust 
grains in the surface brightness maps.
The optical and near-IR surface brightness at the inner
shell ($r/r_{\rm min} \le 1$) was slightly lower in the 
anisotropic case than in the isotropic case, while that 
at the outer shell ($r/r_{\rm min} \ge 1$) was slightly 
higher in the anisotropic case than in the isotropic case.
In general, scattered radiation tends to be brought 
more to the forward direction, i.e., farther away from 
the central star.
Thus, there is more optical to near-IR scattered light 
in the outer shell.
This is corroborated by the slight increase in the 
thermal IR radiation (at 9.8 $\um$) in the shell,
which is caused by the additional dust heating due to 
extra optical light in the outer shell. 

\subsection{Axisymmetric Models\label{axi}}

Having checked the {\twodust} code in spherical cases
under both isotropic and anisotropic scattering 
assumptions,
we have gradually changed the dust density distribution 
from spherical to axial symmetry to observe how the 
change would affect the observable properties of the 
dust shell. 
As a normalized density distribution function for 
the axisymmetric test models, we have adopted a 
so-called ``layered shell model'' that was developed 
to investigate the observed characteristics 
of post-AGB shells \citep{meixner02}:
\begin{equation}\label{axidensfunc}
 \rho (r, \Theta) = 
 \left( \frac{r}{r_{\rm min}} \right)^{%
 -B \left[ 1 + C \sin^{F} \Theta
   \left\{
    \case{e^{-\left( \case{r}{r_{\rm sw}} \right)^{D}}}%
    {e^{-\left( \case{r_{\rm min}}{r_{\rm sw}} \right)^{D}}}
   \right\} 
  \right]} \left[ 1 + A (1 - \cos \Theta)^{F}
   \left\{
    \case{e^{-\left( \case{r}{r_{\rm sw}} \right)^{E}}}%
    {e^{-\left( \case{r_{\rm min}}{r_{\rm sw}} \right)^{E}}}
   \right\}
 \right]. \nonumber   
\end{equation}

Figure \ref{layered} schematically shows the density 
distribution that primarily consists of three layers of 
shells.
The outermost region corresponds to the shell created 
by early AGB mass loss that occurs in almost perfect 
spherical symmetry, and is described by the radial 
fall-off part of the equation ($r^{-B}$).
The axisymmetry arises at two places in the density 
function.
The equatorial enhancement parameter, $A$, introduces 
the overall axisymmetric structure to the shell,
which can be made disk-like or toroidal by the flatness 
parameter, $F$.
The radial fall-off factor, $B$, can additionally be 
a function of the latitudinal angle through the elongation 
parameter, $C$.
These parameters determine the toroidal structure of the 
innermost region of the shell, which is considered to 
be caused by axisymmetric ``superwind'' at the end of 
the AGB phase.

The mid-region of the shell assumes somewhat spheroidal 
dust distribution reflecting the transition of mass loss 
geometry from spherical to axial symmetry during the 
course of the AGB mass loss history.
The symmetry transition parameters, $D$ and $E$, control 
the ``abruptness'' of the transition in the shell: small 
values correspond to slow transition and large values 
correspond to abrupt transition.
The ``superwind'' radius, $r_{\rm sw}$, defines the
``thickness'' of the inner, axisymmetric region of the
shell.

Therefore, the density distribution can be highly 
equatorially enhanced within the superwind radius, 
while it is nearly free of any latitudinal dependence 
at large radii.
Two types of symmetries are thereby described by this 
shell density function.
In the following, we will briefly explore the parameter 
space of the density function to gain some physical 
insights for the behavior of the model results.
Other model parameters are the same as the spherical
shell models with the KMH size distribution.
Table 4 summarizes the parameters for the density function 
used in the axisymmetric test models.

\subsubsection{Equatorial Density Enhancement}

The equatorial enhancement parameter, $A$, sets the 
equator-to-pole density ratio 
($\rho_{\rm eq}/\rho_{\rm pole} = 1 + A$).
Here, we have considered three models:
$A = 0$ and $\tau_{9.8} = 1.0$ (A1, spherical), 
$A = 9$ and $\tau_{9.8} = 1.0$ (A2), and
$A = 9$ and $\tau_{9.8} = 5.0$ (A3).
Figure \ref{axisedA} shows SEDs for these models at
$\thincl = \0deg$ ({\sl gray lines}) and 90$^{\circ}$
({\sl black lines}).
The SED shows the two-peak structure typical of a 
dust-enshrouded system.
The difference in the inclination angle does not affect
the shape of the SED in the A1 models ({\sl solid lines}).
The difference in the visibility of the central star
causes the variation of optical peak flux among 
other cases.

In the A2 and A3 models ({\sl dashed} and {\sl dotted} 
lines, respectively), the equatorially-enhanced dust shell 
can obscure the central star, and the $\thincl = \0deg$
cases ({\sl gray lines}) yield more optical flux than 
the $\thincl = \90deg$ cases ({\sl black lines}).
The A2 model with $\thincl = \90deg$ ({\sl black dashed line}) 
shows more optical flux than the A1 model ({\sl black solid line}).
This is also due to the equatorial enhancement in the dust
density distribution: 
since there is {\sl less\/} dust grains along the polar 
directions in the A2 model than the A1 model, more optical 
light can scatter from the dust shell through such ``bicone 
openings'' of the shell.

The A3 model with $\thincl = \90deg$ ({\sl black dotted line})
displays a highly reddened optical peak due to its optically
thicker shell of $\tau_{9.8} = 5.0$.
This optically thick nature of the A3 model is also seen in the 
mid-IR peak.
The mid-IR peak shows broad emission features at 9.8 and 18.0 $\um$
of amorphous silicates {\sl except for} the $\thincl = \90deg$ 
case of the A3 model, in which the dust column density along the
line of sight is sufficiently high enough to convert the features 
into absorption.

Figures \ref{aximapA1}, \ref{aximapA2}, and \ref{aximapA3} 
show the projected surface brightness maps of the A models 
at three inclination angles.
All the A1 maps (Figure \ref{aximapA1}) appear the same 
irrespective of the inclination angle as expected from a 
spherical model.
However, the A2 (Figure \ref{aximapA2}) and A3 
(Figure \ref{aximapA3}) maps show emission structure that 
is caused by extinction and/or emission due to the 
equatorially-enhanced dust distribution of the shell.
In the A2 model at $\thincl = \90deg$, the optical nebula 
at 0.55 $\um$ shows the classic bipolar shape with the 
central star while the mid-IR nebula at 9.80 $\um$ displays 
two emission peaks characterizing the limb-brightened edges 
of an edge-on dust torus at $\90deg$ inclination.
The orientation of the bipolar lobes in the optical and 
mid-IR are perpendicular to each other.
The toroidal nature of the equatorially-enhanced dust shell 
becomes more apparent in the 9.80 $\um$ maps at smaller 
inclination angles.
At $\45deg$ inclination, only the far side of the torus 
is seen through the bicone opening of the inclined 
torus (which creates the one-sided optical nebula at 0.55
$\um$) as a U-shaped emission peak.
Then, the emission peak shows a complete ring shape 
of the pole-on dust torus at $\0deg$ inclination.

The A3 maps show the optically very thick nature of the 
model.
The central star is completely obscured by the dust lane,
and dust-scattered light creates bipolar lobes extending 
beyond 10{\arcsec} from the central star ($\thincl = \90deg$ 
at 0.55 $\um$).  
The elongated optical emission at 0.55 $\um$ seen at the 
off-center position at $\45deg$ inclination is due to 
scattered emission through the near side of the bicone
opening, and no emission through the far side of the bicone
opening is seen in the map.
The 9.80 $\um$ emission at $\90deg$ inclination shows
two peaks ($\thincl = \90deg$ at 9.80 $\um$).
Unlike the A2 model, the orientation of the mid-IR peaks 
at 9.8 $\um$ is the same as that of the optical bipolar 
lobes at 0.55 $\um$.
Even the mid-IR emission can not escape from the innermost 
region of the shell where the optical thickness is extremely 
high ($\tau_{9.8}=5$).

\subsubsection{Radial Density Fall-Off}

The radial fall-off factor, $B$, is strongly tied to the 
dynamical nature of mass loss provided that the wind velocity 
is constant.
A uniformly expanding shell generated by steady mass loss 
would yield $B = 2$, while a larger $B$ value is expected 
for mass loss with a steadily increasing rate but a smaller 
$B$ value for a diminishing mass loss.
Here, we have considered three cases in which 
$\tau_{9.8}=1.0$: 
$B = 1.5$ (B1; {\sl solid lines}),
$B = 2.0$ (A2; {\sl dashed lines}), and
$B = 3.0$ (B2; {\sl dotted lines}).

SEDs show only a slight difference in the optical peak 
among the models, and the distinction is almost solely 
due to the visibility of the central star through the shell,
i.e., the inclination angle (Figure \ref{axisedB}).
However, there are two major differences in the mid-IR peak: 
the amount of far-IR flux and the strength of the emission 
features.
The difference in the far-IR flux arises because more dust 
is concentrated radially closer to the central star
in models with a large $B$, i.e., a larger amount of 
warmer dust grains is present in the shells having a smaller
$B$ value (hence, {\sl more} far-IR emission).
The B2 models show some far-IR flux difference due to 
inclination.
This stems from the fact that dust distribution in this 
model is highly concentrated to the region close to the 
inner edge of the shell, where the shell has a very flattened 
density distribution: there is only a small column density 
of cold, far-IR emitting dust if the shell is observed at 
pole-on inclination.

There is no major distinction among emission maps of 
these models, except for the very inner emission 
structure at $\90deg$ inclination.
While the B1 maps ({\sl top row} in Figure \ref{aximapB}) 
show the optical bipolar lobes {\sl with} the visible 
central star at 0.55 $\um$ and the limb-brightened mid-IR 
peaks at 9.8 $\um$, the B2 maps ({\sl bottom row} in 
Figure \ref{aximapB}) show the bipolar optical bipolar 
lobes {\sl without} the central star at 0.55 $\um$ and 
a single, O-shaped mid-IR peak at 9.8 $\um$
(the central part has a lower surface brightness than
the O-shaped part).
A highly concentrated dust distribution of the B2 model 
made the central star invisible in the optical, and 
the mid-IR peak more centrally concentrated making it appear
as a single, connected emission peak.

\subsubsection{Shell Elongation}

The $C$ parameter turns on and off the latitudinal dependence 
of the radial fall-off factor, $B$.
Introduction of non-zero $C$ value can slow down the density 
decrease along the polar directions for intermediate
radius ($r \approx r_{\rm sw}$), and therefore, $C$ defines 
the degree of the spheroidal elongation in the transitional 
mid-shell region.
Here, we have considered two cases: $C = 0.5$ (C1) and $C = 3$ (C2).
SEDs in Figure \ref{axisedC} exhibit only a slight 
increase in far-IR emission due to additional dust 
distribution ``filled up'' the bicone openings
of the shell.
The surface brightness maps show the morphological effect 
of shell elongation along the pole.

The C model maps at 90$^{\circ}$ inclination ({\sl left}
frames in Figure \ref{aximapC}) show the morphological 
distinction induced by the parameter $C$.
The $C$ parameter introduces an additional radial fall-off
that can be very steep along the equator.
Thus, the maps are generally more compact if $C$ is
larger, i.e., the C2 maps show more centrally 
concentrated emission.
Moreover, the emission structure of the C2 maps
(especially of optical) are relatively more
elongated along the pole than the C1 maps, giving
an elliptical look to the overall shape of the nebula.
The low-level elliptical elongation is recognizable 
even at the near- and mid-IR wavelengths.

Such a morphology appears to be common in post-AGB 
objects whose shells are optically thin \citep{ueta00}.
The elliptical morphologies of the post-AGB shells 
have been reproduced only by the parameterization of
shell elongation as we introduced in \citet{meixner02}.
In optically thin shells, the amount of emission is 
roughly proportional to the dust column density.
Therefore, an elliptical elongation would suggest
a far slower density drop along the poles than
along the equator, especially because the density 
at the inner region along the equator is much higher 
by default.

This kind of density distribution requires a peculiar 
mass loss.
After the initial spherically symmetric mass loss, 
enhancement of mass loss first occurs in the polar
directions, making the shell elliptically elongated.
Then, the rate of mass loss into the equatorial directions
starts to increase, and it keeps increasing until the 
equatorial mass loss rates surpass the polar mass loss 
rates.
Once this happens, the equatorially enhanced structure 
can ensue in the subsequent evolution of mass loss geometry.
Further investigations of this peculiar mass loss
history is crucial to identify and understand mechanisms 
to generate shells that are toroidal in the interior
but elliptical in the surrounding regions.

\subsubsection{Symmetry Transition in the Shell}

The $D$ and $E$ parameters describe the ``abruptness'' of 
the geometrical transition, which has to depend on the
physical nature of the emergence of axisymmetry.
Larger $D$ and $E$ parameters indicate more abrupt 
dissipation of the latitudinal variation in the 
density distribution.
By independently adjusting parameters $D$ and $E$, 
one can control which of the two latitudinal dependence 
of the dust density distribution (on $A$ or $B$) would 
persist at large radii.
Here, we have considered four $\tau_{9.8} = 5.0$ cases,
in which
$D = 1$ and $E = 1$ (model D1), 
$D = 3$ and $E = 3$ (model D2), 
$D = 3$ and $E = 1$ (model E1), and 
$D = 1$ and $E = 3$ (model E2).

SEDs do not show any significant difference among models
(Figure \ref{axisedD}).
The difference in the inclination is seen as a presence 
or absence of the silicate absorption features at 9.8 and 
18.0 $\um$ on the mid-IR peak and as the discrepancy in the 
optical peak arising from the visibility of the central star.
Although the $\90deg$ cases ({\sl black lines}) do not show 
much distinction, the $\0deg$ cases ({\sl gray lines}) are 
different in the amount of optical emission.
This is because the column density along the equatorial
plane is almost uniformly set by the input value of 
$\tau_{9.8} = 5.0$ 
among the models, whereas the column density along the 
pole differs depending on the parameters $D$ and $E$.
Dust density at $\rmin$ along the equator 
($\rho_{\rm min}$) is set by $\tau_{9.8}$, while that 
along the pole is scaled from $\rho_{\rm min}$ through 
eq.\ (\ref{axidensfunc}).
Since $\rho_{\rm min}$ becomes large when $E$ is large,
dust density at $\rmin$ along the pole is larger for 
the D2 and E2 models than for the D1 and E1 models.
Thus, the D2 and E2 models suffer from more optical 
self-extinction than the D1 and E1 models.
The detailed radial profile along the pole is determined
by the specific choice of $D$, and models with a smaller 
$D$ value would have a slower density fall-off (i.e., 
a larger column density
along the pole).
Hence, the D1 model suffers from more self-extinction 
than the E1 model, and the E2 model suffers from more 
self-extinction than the D2 model.
Therefore, the E2 model sees the largest self-extinction 
in the optical, followed by the D2, D1, and E1 models.
This effect is also seen in the $\90deg$ cases
but at a much lower level. 

Figure \ref{aximapDE} shows the brightness maps of the
models D1, D2, E1 and E2 at $\90deg$ inclination
({\sl first}, {\sl second}, {\sl third}, and 
{\sl fourth} row, respectively).
Maps at other inclination angles are generally similar 
to the A3 maps (Figure \ref{aximapA3}).
The effect of these parameters on the surface brightness 
morphology becomes pronounced when $D$ and $E$ have different 
values.
The density structure of the E1 model ($D > E$) consists of a
generally more gentle rise in the outer region and a steeper 
rise in the innermost region along the equator than along
the pole.
Thus, the high emission bipolar lobes are very elongated along the
pole (i.e., more scattered light towards the bicone openings)
while a low emission nebula is elongated along the equator 
(i.e., a wider large dust lane between the optical lobes
and the oblate shape of the near-IR emission).
On the contrary, the density structure of the E2 model ($D < E$)
has a steeper rise in the outer region and a more gentle 
rise in the innermost region along the equator than along
the pole.
This results in a rather flattened optical bipolar lobes 
along the pole
(i.e., the polar extent of the optical lobes is the smallest)
whose emission level precipitously drops at far radii along 
the equatorial plane.
The near-IR emission map of the E2 model shows the elliptical
elongation caused by the effect of slower density fall-off at 
large radii along the poles.
Hence, the $D$ and $E$ parameters mainly influence low-level 
emission arising from dust scattering, and therefore do not 
seem to affect the mid-IR morphologies.
		
\subsubsection{Flatness of the Equatorial Enhancement}

The $F$ parameter sets the ``flatness'' of the equatorially 
enhanced density distribution of the shell.
Small $F$ values yield toroidal density distributions,
while large $F$ values result in disk-like structures.
Here, we present two cases with $F=3$ (model F1) and
$F=9$ (model F2) at $\tau_{9.8} = 1.0$.
Model SEDs (Figure \ref{axisedF}) show very little 
difference except for the optical emission.
The $F=9$ cases put out more optical emission than 
the $F=3$ cases, and this is because more optical light 
can be scattered through the shell when the density 
distribution is more flattened.
However, the F1 model puts out more emission in the 
redward of the near-IR than the F2 model, simply because 
the amount of thermal IR emission is directly proportional 
to the amount of absorbed optical radiation.
The distinction due to the inclination is again a result of
the visibility of the central star.

The morphological differences appear most obvious in the 
edge-on (90$^{\circ}$) surface brightness maps.
The $F=9$ models (Figure \ref{aximapF2}) show more extension
along the equatorial direction and less extinction along the
polar directions with respect to the $F=3$ models 
(Figure \ref{aximapF1}).
The distinction is visible even in the near-IR emission maps.
However, a highly flattened density distribution alternatively 
means lower column density when observed closed to pole-on.
Thus, $\45deg$ cases of the F2 model show less elongation than
those of the F1 model.

\subsection{Anisotropic Scattering in Axisymmetric Shells}

The above axisymmetric cases are all done under the 
assumption of isotropic scattering.
Here, we allow anisotropic scattering by dust grains.
To observe the effects of different ways of dust scattering,
we used a model for a post-AGB star, \iras 17150$-$3224 
\citep{meixner02}.
This model uses the KMH size distribution with 
$a_{0} = 200$ $\um$.
Large grains tend to scatter more to the forward direction, 
and thus, scattered radiation tends to go farther along 
the direction of incident rather than sideways.
Therefore, we would expect to see more extended reflection
nebulosities.
In this particular case, 
the bipolar lobes are expected to appear farther away in 
the anisotropic scattering case than in the isotropic 
scattering case.

Figure \ref{axianisosed} shows SEDs for the cases
in which both isotropic ({\sl solid line}) and anisotropic 
({\sl dashed line}) scattering are considered.
The anisotropic scattering assumption yielded more 
scattered radiation in the optical and near-IR wavelength 
($< 6$ $\um$).
This is interpreted as a larger amount of optical to 
near-IR radiation is being scattered out of the optically 
thick dust torus through the bicone openings of the torus.
Meanwhile, there is no change in the absorptivity of the 
grains and the IR excess remains the same between the model
calculations.

Figure \ref{axianisoslice} show the cross-cuts of the 
surface brightness at different inclinations
along the major axis of the shell 
(see Figures 5 and 7 of \citet{meixner02} for
the 2-D projected images of the model).
The equatorial density enhancement of the shell is so 
high ($1+A = 160$) that scattered radiation can escape
only through the bicone openings of the dust torus.
At the pole-on orientation (0$^{\circ}$), 
radiation into the equatorial region tends to be brought
farther into the direction of radiation, 
and hence, less emission gets scattered towards the 
observer via $\90deg$ scattering resulting in a narrower 
profile in the anisotropic case.
Thus, reflection nebulosities appear smaller in the 
anisotropic case than in the isotropic scattering case
at small inclination angles.

The surface brightness profiles show a markedly
distinct behavior at inclination larger than $\45deg$.
There is less emission in the equatorial region (near
the central star, i.e., in the dust lane; 
closer than about $0\farcs2$ from the center) but there
is more emission farther away in the bipolar lobes
(farther than about $0\farcs2$ from the center)
in the anisotropic case than in the isotropic case.
In effect, the lobes appear farther away from each other
in the anisotropic case than in the isotropic case,
as expected from a simple argument using a tendency
towards forward scattering in the anisotropic cases.

\subsection{Optical Depth versus Inclination Angle}

The ubiquity of dusty shells/disks prompts the need
for a simple ``signature'' to interpret the geometry
(e.g., the degree of flattening or inclination) of
the shell/disk from the shape of the SED.
With a 2-D code like {\twodust}, we can observe how
SEDs change depending on the inclination angle
and the geometry of the shells/disks.
Thus, we took the \iras 17150$-$3224 model 
\citep{meixner02}, which has a highly flattened dust 
shell (i.e., the pole-to-equator density ratio of 160), 
and varied its inclination angle, $\theta_{\rm incl}$,
and equator-to-pole density ratio, 
$\rho_{\rm eq}/\rho_{\rm pole}$.
Figure \ref{17150} shows the results of this 
exercise.

Figure \ref{17150}a ({\sl top frame}) shows the SEDs 
at different inclination angles.
As $\theta_{\rm incl}$ increases, the amount of optical 
light is greatly reduced in accordance with the 
decreasing visibility of the central star through the 
``bicone opening'' of the dust torus, while the mid-IR 
peak does not display a significant change except for 
the increasing depth of the 9.8 $\mu$m silicate absorption 
feature.
The slight decrease of the mid-IR emission is due to
self-extinction by the dust torus induced by the
inclination of the system.
Figure \ref{17150}b ({\sl bottom frame}) shows the 
SEDs of the shell with three different 
$\rho_{\rm eq}/\rho_{\rm pole}$ (160, 100, and 50) 
at both pole-on 
($0^{\circ}$) and edge-on ($90^{\circ}$) orientations.
In the edge-on cases, there is no
apparent shift in the SED shape, since the optical
light from the central star is completely obscured
irrespective of the value of $\rho_{\rm eq}/\rho_{\rm pole}$
with the given optical depth of the shell.
In the pole-on cases, the amount of optical light 
decreases as $\rho_{\rm eq}/\rho_{\rm pole}$ decreases.
This occurs because there is more dust along the poles
when $\rho_{\rm eq}/\rho_{\rm pole}$ is low and hence
there is more absorption of starlight by the dust 
grains.

For a given model, the shape of the SED can be 
easily understood by considering the energetics 
of the dust shell appropriate for that particular 
model.
However, it is difficult to figure out the shell 
geometry from the SED shape alone, since a particular
SED shape can be generated by models having
different inclination and geometry.
For example, almost identical SEDs are generated
by the $30^{\circ}$ inclination model
({\sl dotted line} in Figure \ref{17150}a) and
the $0^{\circ}$ inclination but 
$\rho_{\rm eq}/\rho_{\rm pole} = 50$ model
({\sl dash-dotted line} in Figure \ref{17150}b),
let alone
the $90^{\circ}$ inclination models having
distinct $\rho_{\rm eq}/\rho_{\rm pole}$ values
(Figure \ref{17150}b).
The major difference in the SED shape at a given 
optical depth is the amount of optical light,
as we have seen.
In reality, the interstellar extinction can 
cause an additional reduction of the optical peak,
and SEDs of geometrically distinct shells
can appear very similar.

SEDs give only the total amount of radiation at
given wavelengths, which is surface brightness 
integrated over the entire spatial extent of the 
shell.
However, the surface brightness at a particular 
region of the shell is related to the 
optical depth/self-extinction along the line of 
sight towards that particular region of the shell,
which is strongly dependent on the geometry and
the inclination of the shell.
Different geometry and inclination angles alter
the optical depth/self-extinction at different
regions in the shell.
Once surface brightnesses are integrated to yield 
flux, we lose all the spatial information of
the local optical depth within the shell that is 
necessary to decipher the energetics of the shell.
Therefore, it is intrinsically difficult to
recover geometric information from spatially 
unresolved SEDs.

High-resolution images, especially at mid-IR
wavelengths, in combination with an SED provide 
excellent constraints on the geometry of the 
dusty shells/disks.
The 2-D projected images are strongly dependent 
on the self-extinction of the shell caused by 
the geometry and inclination angle.
Therefore, the direct probe of dust distribution
at the mid-IR wavelengths plays a critical role
in determining the dust energetics and constraining
the model parameters (e.g., 
\citealt{ueta01a,ueta01b,meixner02}).

\section{Summary\label{2dustconcl}}

In order to numerically model a dust-enshrouded axisymmetric 
system, we have developed a dust radiative transfer code, 
{\twodust}.
This code solves the equation of radiative transfer in a 
2-D polar grid by considering a fully 3-D radiation field.
In this sense, the {\twodust} code is a 2.5 dimensional 
radiative transfer code.
A solution is sought through an iterative scheme 
originally developed by \citet{collison91}, in which 
self-consistency of the solution is achieved 
by requiring a global luminosity constancy at each 
radial grid.

The converged solution constrains the intensity and 
dust temperature fields, from which we derive observables 
such as SED and projected surface brightness maps
at given inclination angles.
Dust grains in the axisymmetric system are considered 
to be the only means of radiative energy transport.
Thus, proper considerations of the optical properties 
and size distributions of dust grains are of particular 
importance in our analysis.
To calculate the absorption and scattering cross sections
of dust grains, we use Mie theory supplied with 
laboratory-measured refractive indices of ``real'' dust 
grains.

We have first tested the code for a spherically symmetric 
shell, comparing the {\twodust} results with the results 
generated by a popular 1-D radiative transfer code, 
{\sf DUSTY}.
The results obtained by these codes agreed quite well.
Then, the scattering assumption is extended
from a simple isotropic case to an anisotropic case.
As expected in a spherically symmetric dust distribution, 
no major difference between isotropic and anisotropic
cases has been recognized.

Axisymmetric models are then tested by gradually departing 
from spherical symmetry. 
We have employed a specific dust density distribution 
function that resulted in our investigation of post-AGB 
shell morphologies.
Although our exploration in the parameter space is far
from thorough, the results of the test models have 
demonstrated the relationship between the model parameters 
and the resulting observables (among all, the projected 
shell morphologies).
As in the spherical cases, the assumptions of isotropic 
and anisotropic scattering are both tested with the 
axisymmetric models.
The test results are consistent with the expectations
that the presence of large grains would bring the 
anisotropically scattered light farther away from the 
source of scattered radiation based on the known 
forward/backward scattering tendency of dust grains.

These tests have demonstrated the basic capabilities 
of this new code fairly well.
No spurious result has been generated in the entire
test model runs, and therefore, we are reasonably
confident to conclude that the {\twodust} code 
would produce good dust radiative transfer 
models for an axisymmetric system.
With these model runs, we have also been able to
characterize the effects of various parameters 
involved in the model to the model results, 
especially the resulting SED and shell morphologies.
Although such characterizations are far from complete,
this is nonetheless a necessary and important first 
step to confidently apply the {\twodust} code in the 
following model fitting using observed data.
In fact, the {\twodust} code has been applied to
model the post-AGB shells and have successfully 
produced the best models to date to suggest that
the two distinct types of post-AGB shell morphologies
arise mainly from the difference in the optical
depth of the shells \citet{meixner02}.

We also demonstrated the difficulty of constraining
the shell geometry and inclination angle solely from
the shape of the SED, since spatially unresolved SEDs
do not provide any spatial information necessary to
constrain the geometry and inclination angle of the 
dusty shells/disks.
Such spatial information needs to be obtained from
high-resolution images.
To investigate the energetics and geometry of the 
dusty shells/disks, especially important are mid-IR
images that directly probes the dust distribution
within the shells/disks.

\acknowledgments
Ueta \& Meixner have been supported by NSF CAREER Award grant,
AST 97-33697.
An anonymous referee is thanked for valuable comments.
The {\twodust} code is a public code, and a distribution 
package can be obtained by contacting the authors.

\clearpage

\begin{deluxetable}{ll}
\tablecolumns{2} 
\tablewidth{0pt} 
\tabletypesize{\small}
\tablecaption{Summary of Input Parameters for \protect{\twodust} 
Models\label{inputs}} 
\tablehead{%
\colhead{Parameter} & 
\colhead{Description}} 
\startdata 
\multicolumn{2}{c}{Grid}\\[1.0ex]
\hline
$n_{r}$         & number of radial grid points  \\
$n_{\Theta}$    & number of latitudinal grid points \\
$n_{\theta}$    & number of $\theta$ directional grid points \\
$n_{\phi}$      & number of $\phi$ directional grid points \\
$n_{\theta}^{(m)}(r)$& number of grid points in the $\theta^{(m)}$ zone 
				  ($m = 1$, 2, and 3) \\
$\theta^{(2)}(r)$  & angle of the $\theta^{(2)}$ boundary \\
{\sl MAXSTEP}   & maximum number of line integration steps \\
{\sl VSPACE}    & line integration step size factor, $\beta$ \\
\cutinhead{Central Star and Shell}
$\rstar$ ($\rsun$) & radius of the central star \\
$\teff$ (K)     & effective temperature of the central star \\
$d$ (kpc)       & distance to the system \\
$v_{\rm exp}$ (km s$^{-1}$)  & expansion velocity of the shell \\
$r_{\rm min}$ (arcsec) & inner shell radius \\
$r_{\rm max}$ ($r_{\rm min}$) & outer shell radius \\
$r_{\rm sw}$ ($r_{\rm min}$)  & superwind shell radius (when defined) \\
$\tau_{0}$ & optical depth along the line of sight directly to the central star \\
$\lambda_{\tau}$ & wavelength at which $\tau_{0}$ is defined (must be in the $\lambda$ grid) \\
$n_{\rm layer}$ & number of composition layers in the shell \\
$r_{\rm layer}(i)$ ($r_{\rm min}$)& location of the composition boundaries 
                  ($i = 1, \cdots, n_{\rm layer}-1$) \\
$A - F$         & six parameters reserved for the density function \\
\cutinhead{Dust Grains}
$n_{\lambda}$  & wavelength ($\lambda$ in $\mu$m) grid points \\
$n(\lambda)$, $k(\lambda)$ & complex refractive index ($m = n+i k$) 
as a function of $\lambda$\\ 
$\rho_{\rm bulk}$ (g cm$^{-3}$)& bulk density of the dust species \\
$\gamma$        & exponential factor for the size distribution function \\
$a_{\rm min}$ ($\mu$m)  & minimum grain size \\
$a_{\rm max}$ or $a_{0}$ ($\mu$m) & maximum grain size or grain size scale factor \\
{\sl AFLAG} & choice of dust size distribution (MRN or KMH) \\
{\sl DFLAG} & choice of the number or mass density constancy in the density function\\
{\sl SFLAG} & choice of the scattering mode (isotropic or anisotropic) 
\enddata
\end{deluxetable}

\begin{deluxetable}{ll}
\tablecolumns{2} 
\tablewidth{0pt} 
\tabletypesize{\small}
\tablecaption{Summary of Output Parameters for 
\protect{\twodust} Models\label{outputs}} 
\tablehead{%
\colhead{Parameter} & 
\colhead{Description}} 
\startdata 
$J_{\nu}(r$, $\Theta$) (erg s$^{-1}$ cm$^{-2}$ Hz$^{-1}$ sr$^{-1}$)& 
mean specific intensity (when isotropic)\\
$I_{\nu}(r$, $\Theta$; $\theta$, $\phi$) (erg s$^{-1}$ cm$^{-2}$ Hz$^{-1}$ sr$^{-1}$)& specific intensity (when anisotropic)\\
$T(r$, $\Theta$) (K)& dust temperature \\
$F_{\nu}$ (Jy) & total specific flux density (SED) \\ 
$I_{\nu}$ (Jy arcsec$^{-2}$) & surface brightness (2-D projected map) \\
$\tau_{\lambda}$ & optical depth (2-D projected map) \\
$\rho (r_{\rm min}, 0)$ (g cm$^{-3}$) & mass density at inner radius on the pole \\
$\rho (r_{\rm min}, \case{\pi}{2})$ (g cm$^{-3}$) & mass density at inner radius on the equator\\
$M_{\rm AGB}$ ($M_{\odot}$) & total mass of dust in the AGB shell \\ 
$M_{\rm sw}$ ($M_{\odot}$) & total mass of dust in the superwind shell \\
$t_{\rm dyn}^{\rm AGB}$ (yr) & timescale for AGB mass loss \\
$t_{\rm dyn}^{\rm sw}$ (yr) & timescale for superwind mass loss \\
${\dot M}_{\rm AGB}$ ($M_{\odot}$ yr$^{-1}$) & AGB dust mass loss rate \\
${\dot M}_{\rm sw}$ ($M_\odot$ yr$^{-1}$)& superwind dust mass loss rate\\
$\kappa_{\lambda}$, $\sigma_{\lambda}$ & absorption and scattering 
cross sections \\
\enddata
\end{deluxetable}

\begin{deluxetable}{ll}
\tablecolumns{2} 
\tablewidth{0pt} 
\tabletypesize{\small}
\tablecaption{Spherical Shell Model Parameters\label{twodust_tab1}} 
\tablehead{%
\colhead{Parameter} & 
\colhead{Value}} 
\startdata 
\multicolumn{2}{c}{Grid}\\[1.0ex]
\hline
$n_{r}$ & 45 \\
$n_{\Theta}$ & 8 \\
$n_{\theta}$ & $11 - 16$ \\
$n_{\phi}$ & 8 \\
$n_{\lambda}$ & 30 \\
\cutinhead{Central Star}
$\rstar$ (cm) & $2.38 \times 10^{12}$ \\
$\teff$ (K) & 7000 \\
$d$ (kpc) & 1.0 \\
$v_{\rm exp}$ (km s$^{-1}$) & 10.0 \\
\cutinhead{Shell}
$r_{\rm min}$ (cm) & $1.125 \times 10^{16}$ \\
$r_{\rm max}$ (cm) & $56.25 \times 10^{16}$ \\
\cutinhead{Dust Grains}
$\rho_{\rm bulk}$ (g cm$^{-3}$) & 3.71\phn \\
MRN \\
~~~$a_{\rm min}$ ($\um$) & 0.005 \\
~~~$a_{\rm max}$ ($\um$) & 0.25\phn \\
~~~$\tau_{9.8 \um}$ & 0.5, 1.0, 5.0 \\ 
KMH \\
~~~$a_{\rm min}$ ($\um$) & 0.005 \\
~~~$a_{0}$ ($\um$) & 0.20\phn \\
~~~$\tau_{9.8 \um}$ & 0.5, 1.0, 5.0 \\
\cutinhead{Execution}
CPU time & a few hours to days \\
&  (on Sun Blade 100) \\
\# of Iteration & 5 to 8 times
\enddata
\end{deluxetable}

\begin{deluxetable}{lccccccc}
\tablecolumns{8} 
\tablewidth{0pt} 
\tablecaption{Axisymmetric Model Density Function Parameters\label{aximodparam}} 
\tablehead{%
\colhead{Model} & 
\colhead{$A$} &
\colhead{$B$} &
\colhead{$C$} &
\colhead{$D$} &
\colhead{$E$} &
\colhead{$F$} &
\colhead{$\tau_{9.8 \um}$}}
\startdata 
A1 & 0 & 2.0 & 0.0 & 0 & 0 & 1 & 1.0 \\
A2 & 9 & 2.0 & 0.0 & 0 & 0 & 1 & 1.0 \\
A3 & 9 & 2.0 & 0.0 & 0 & 0 & 1 & 5.0 \\
B1 & 9 & 1.5 & 0.0 & 0 & 0 & 1 & 1.0 \\
B2 & 9 & 3.0 & 0.0 & 0 & 0 & 1 & 1.0 \\
C1 & 9 & 2.0 & 0.5 & 3 & 3 & 1 & 1.0 \\
C2 & 9 & 2.0 & 3.0 & 3 & 3 & 1 & 1.0 \\
D1 & 9 & 2.0 & 1.0 & 1 & 1 & 1 & 5.0 \\
D2 & 9 & 2.0 & 1.0 & 3 & 3 & 1 & 5.0 \\
E1 & 9 & 2.0 & 1.0 & 3 & 1 & 1 & 5.0 \\
E2 & 9 & 2.0 & 1.0 & 1 & 3 & 1 & 5.0 \\
F1 & 9 & 2.0 & 1.0 & 3 & 3 & 3 & 1.0 \\
F2 & 9 & 2.0 & 1.0 & 3 & 3 & 9 & 1.0 \\
\enddata
\end{deluxetable} 

\clearpage

\appendix

\section{Method of Computation\label{detailmethod}}

\subsection{Axisymmetric Computational Grid}

In {\twodust}, the dusty shell is assumed to extend from 
the inner radius, $\rmin$, to the outer radius, $\rmax$, 
around a central source.
While the presence of gas in the shell is neglected,
dust and gas are assumed to be well-coupled and 
well-thermalized.
The grid points are defined using spherical polar 
coordinates, and axial symmetry is assumed by default:
the dust density profile is expressed by
a 2-D function, $\rho (r, \Theta)$.
We also assume symmetry with respect to the equatorial 
plane of the system.
However, we use a grid with full $\pi$ radians in the 
$\Theta$ direction to compute physical quantities by 
full 3-D ray tracing (see \S \ref{rad}).

The computational grid consists of $\nrad$ radial 
zones centered at
\begin{equation}
  r_{i} = 
  \rmin e^{\gamma \left( i - \case{1}{2} \right)^{2}} 
  \quad \mbox{where} \quad
  \gamma = 
  \frac{1}{\nrad^{2}} \ln \left(\frac{\rmax}{\rmin}\right)
  \quad \mbox{for} \quad i=1, \cdots, \nrad
\end{equation}
and of $\nlat$ latitudinal zones centered at the 
Gaussian-Legendre quadrature points. 
The temperature of dust grains ($\tdust$ in units of K) 
and the specific intensity ($I_{\nu}$ in units of erg 
s$^{-1}$ cm$^{-2}$ Hz$^{-1}$ sr$^{-1}$) of the radiation 
field within the shell are determined at each zone center.

\begin{figure}[h]
\begin{center}
\hfill
 \includegraphics[width=2.8in]{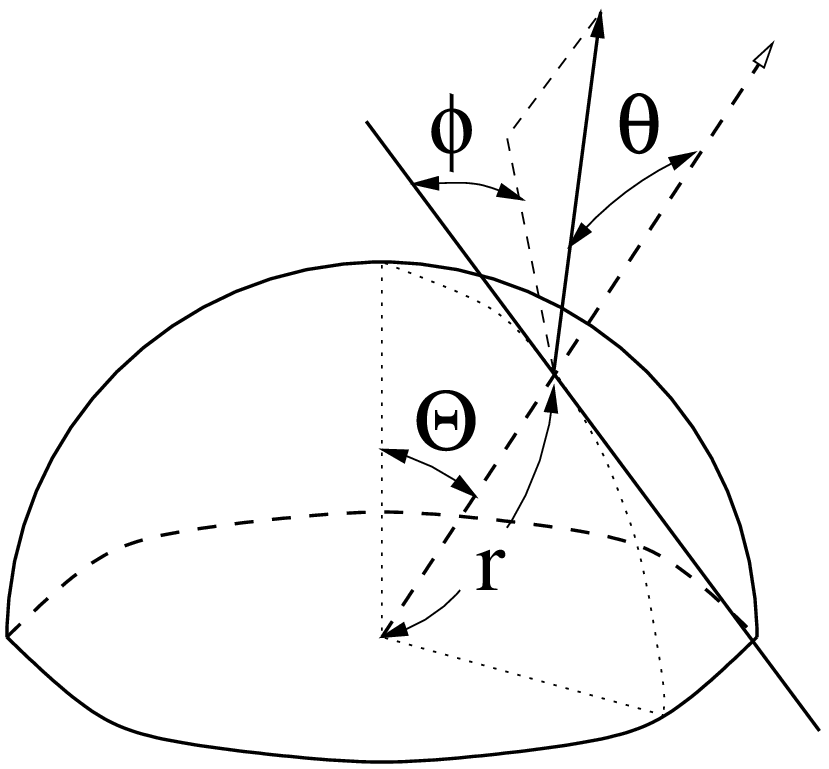}
\hfill
 \includegraphics[width=3.5in]{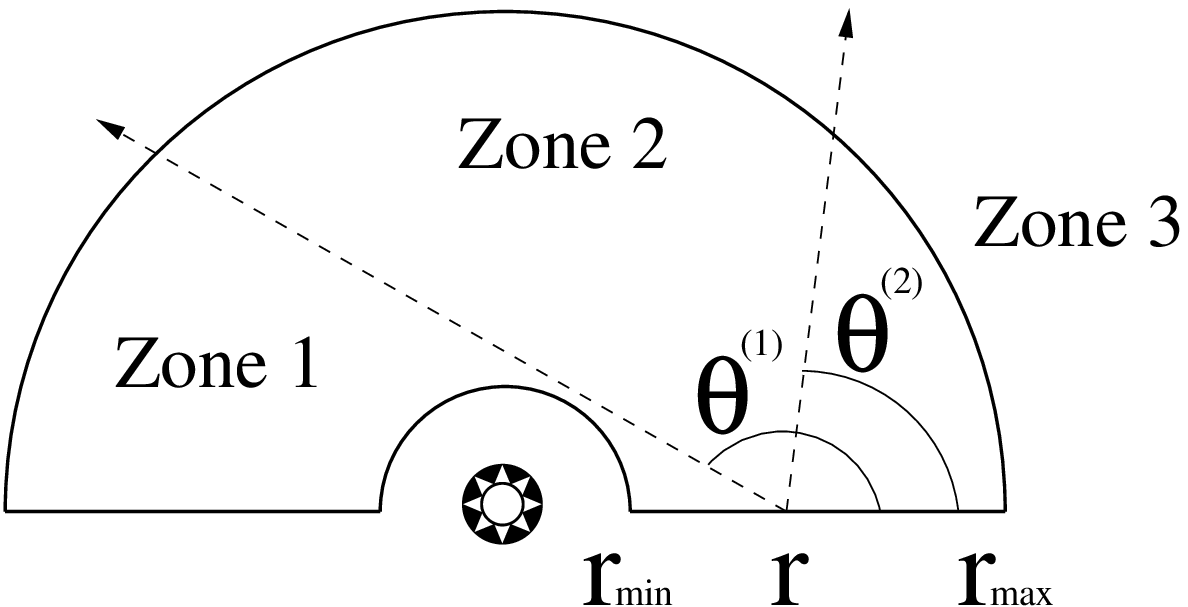}
\hfill
\end{center}
 \figcaption{\label{2ddirection}%
 [{\sl Left}]
 Definition of the position, ($r, \Theta$), in the shell 
 and the directions, ($\theta, \phi$).
 [{\sl Right}]
 Schematic representation of the dust shell and the 
 definitions of the $\theta$ directions.
 At the present grid location, $(r, \Theta)$, the $\theta$ 
 zone boundaries are at $\theta^{(1)}$ and $\theta^{(2)}$
 from the outward radial direction.
 There are $n_{\theta}^{(1)}$, $n_{\theta}^{(2)}$, and 
 $n_{\theta}^{(3)}$ Gaussian quadrature spaced directions 
 respectively in each of the three $\theta$ zones.}
\end{figure}

At each grid, $(r_{i}, \Theta_{j})$, the directions are 
defined by the angles $\theta$ and $\phi$, where $\theta$ 
is the angle measured {\sl from} the radially 
outward direction and $\phi$ is the angle in the azimuthal 
direction with respect to the radially outward direction 
measured from the plane of reference defined by the radially 
outward vector and the pole (z-axis) of the system
(Figure \ref{2ddirection}).
For the $\theta$ direction, we set three ``zones'' over
the $0 \le \theta \le \pi$ range, in each of which 
$n_{\theta}^{(m)}$ directions (the superscript, $m$, 
indicates the zone running from 1 to 3) are defined 
in order to efficiently sample the dust shell (Figure 
\ref{2ddirection}).
The first $\theta$ zone (zone 1) is defined to subtend 
the inner cavity ($r \le \rmin$) of the shell.
The second $\theta$ zone (zone 2) covers the region where 
the dust shell is the brightest, i.e., from which most of 
the radiation is expected to arise.
The third $\theta$ zone (zone 3) covers ``the rest'' 
of the $\theta$ angles, from which not much radiation is 
expected in general.
Thus, there are 
$n_{\theta}^{(1)} + n_{\theta}^{(2)} + n_{\theta}^{(3)}$
angles in the $\theta$ direction.
The size of the three zones would normally be different 
at different radial location, and the number of directions
in each of the $\theta$ zones ($n_{\theta}^{(1)}$, $n_{\theta}^{(2)}$, 
and $n_{\theta}^{(3)}$) also need to be defined 
depending on the structure of the shell to effectively
sample the dust shell.
For the $\phi$ direction, there are $\nphi$ directions 
for the range of $0 \le \phi \le \pi$, 
and symmetry is assumed with respect to the plane of 
reference defined above, i.e., there are 2 $\nphi$ 
discrete directions defined for the full $2 \pi$ radians
in the $\phi$ direction.

Therefore, if we set
$\ntheta=n_{\theta}^{(1)} + n_{\theta}^{(2)} + n_{\theta}^{(3)} = 16$
and $\nphi=8$ at a particular grid point, there would be
the total of 256 directions (rays) defined at this grid point.
Along each of these rays, the equation of radiative 
transfer needs to be solved to compute the amount of radiation
available at this particular grid point.

\subsection{Radiative Transfer\label{rad}}

In {\twodust}, the specific intensity, 
$I_{\nu} (r, \Theta; \theta, \phi)$, is 
derived using the formal solution of the radiative 
transfer equation,
\begin{equation}\label{transfer}
 I_{\nu} (r, \Theta; \theta, \phi) =
  \int_{(r, \Theta)}^{(r_{0}, \Theta_{0})}
  S_{\nu} (r^{\prime}, \Theta^{\prime}; \theta, \phi) 
  e^{-\tau_{\nu} (r^{\prime}, \Theta^{\prime})} 
  d \tau_{\nu},
\end{equation}
where $S_{\nu}$ is the source function
and $\tau_{\nu}$ is the optical depth along a particular 
ray.
This line integration is carried out from the given point 
$(r, \Theta)$ to the point $(r_{0}, \Theta_{0})$ where the 
shell ends or the local $S_{\nu} (r^{\prime}, \Theta^{\prime})$ 
becomes too small to contribute to $I_{\nu}$ (hence, the
{\sl long} characteristic scheme).
The line integration must be done for each of the rays 
at each grid point.

The step size of the line integration, $d \tau_{\nu}$, 
is related to the physical step size, $dl$, by
$d \tau_{\nu} = \kappa_{\nu} \rho d l$.
To allow enough sampling along a characteristic, we set 
$dl$ to be {\sl much} smaller than the local minimum 
photon mean free path, i.e.,
$\lambda_{\rm min}^{\rm mfp} \gg d l = \beta / (\kappa \rho)$,
where $\beta$ is a scaling factor that is much less than 
unity.
To minimize $dl$, we use the largest $\kappa$, which is 
typically $\kappa$ at the largest $\nu$.
Then, the step size depends only on the local density 
of the shell and is not dependent on the local radiation 
field.
Hence, if we determine the size of all the line 
integration steps along all rays once and for all for 
the given $\rho (r, \Theta)$ before the iterative 
processes of radiative transfer begin, we do not need 
to recalculate them for the rest of the iterative 
processes.

The {\twodust} code thus generates a ``template'' 
of the line integration consisting of the total number 
of line integration steps, and each step size for all 
the characteristic defined at each of the $n_{r}$ grid 
points {\sl on the equator} since the density is the 
{\sl highest} along the equator and then rays are always 
finely sampled near the grid points.
Then, this tailor-made ``line integral template'' for
a particular density distribution at hand is used during 
the entire duration of the iterative processes of radiative 
transfer calculations, and hence, the entire computation 
time is spent to find a converged solution.

After completing line integration for all characteristics,
we angle-average the sum to derive the mean specific intensity,
$J_{\nu}$, under the isotropic scattering assumption.
The local temperature of dust grains is then determined
by assuming radiative equilibrium between radiation and
dust grains through
\begin{equation}\label{radeq}
 \int_{0}^{\infty} \kappa_{\nu} 
  B_{\nu} \left( T (r, \Theta) \right) d \nu
  = \int_{0}^{\infty} \kappa_{\nu} 
  J_{\nu} (r, \Theta)
  d \nu.
\end{equation}
Here, $\kappa_{\nu}$ is the absorption cross section
and $B_{\nu}$ is the Planck function.
For a given $J_{\nu}$, we immediately obtain 
$\kappa_{\nu} B_{\nu}$,
from which we deduce the value of $T (r, \Theta)$.

To constrain the radiation and temperature fields 
self-consistently, the {\twodust} code follows the 
iterative method elucidated by \citet{collison91} that
is briefly illustrated below.
At each iterative step, the values of 
$I_{\nu} (r, \Theta; \theta, \phi)$ 
are needed to calculate the source function.
Then, new values of 
$I_{\nu} (r, \Theta; \theta, \phi)$ 
and $T (r, \Theta)$ are derived through 
equations (\ref{transfer}) and (\ref{radeq}).
These new values are different from the original 
values, and thus this process must be repeated 
until a converged solution is found.
We maintain self-consistency by requiring the total 
luminosity at each radial grid be equal to the stellar 
luminosity, $\lstar$.
This leads to a recursive relation that is used to 
update the values of 
$I_{\nu} (r, \Theta; \theta, \phi)$ and $T (r, \Theta)$,
and hence, the source function for the next iteration step.
The iteration is repeated until the luminosity constancy 
is achieved within a desired limit at each radial grid.
If the current and previous values of 
$I_{\nu}  (r, \Theta; \theta, \phi)$ are equal when the 
desired condition is met, we have a self-consistent solution 
to our radiative transfer problem.

\section{Dust Properties}

With the laboratory-measured complex refractive index 
($m = n + i k$), we can calculate the ``$Q$'' efficiency 
factors for the dust cross sections using Mie theory 
\citep{hulst57,bohren83}.
For a spherical particle of radius, $a$, 
the absorption and scattering cross sections, 
$\kappa_{\nu}$ and $\sigma_{\nu}$, at a 
particular frequency are calculated by
\begin{equation}\label{cross}
  \left( \begin{array}{l}
   \kappa_{\nu} \\
   \sigma_{\nu}
   \end{array} \right)  = \sum_{i}
  \alpha_i
 \int_{a}
 \left( \begin{array}{l}
  Q_{i}^{\rm abs} (a, \nu) \\
  Q_{i}^{\rm sca} (a, \nu)
 \end{array} \right)
 \pi a^2 n_{i}(a) da,
\end{equation}
where the subscript $i$ refers to the $i$-th dust species
in the shell.
Here, $Q_{i}^{\rm abs} (a, \nu)$ and $Q_{i}^{\rm sca} (a, \nu)$ 
are the size- and frequency-dependent ``$Q$'' factors 
obtained from Mie theory and 
$\alpha_i$ is the weighting factor based on the abundance of 
the specific dust species.
$n_{i}(a)$ is the normalized dust size distribution function
of the species, $i$, 
for which we use one of the following two forms;
\[ n(a) \propto \left\{ \begin{array}
		{l@{\quad:\quad}ll}
		 a^{-\gamma} & 
		 a_{\rm min} \le a \le a_{\rm max} &
		 {\rm (MRN)} \\
		 a^{-\gamma} e^{-a/a_{0}}  &
		  a_{\rm min} \le a & 
		  {\rm (KMH)}.
		  \end{array}  \right. \]

\begin{figure}[h]
\begin{center}
\hfill
 \includegraphics[width=3.5in]{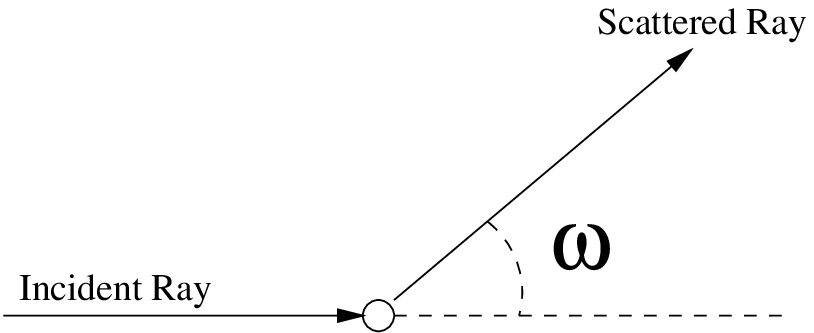} 
\hfill
 \includegraphics[width=2.5in]{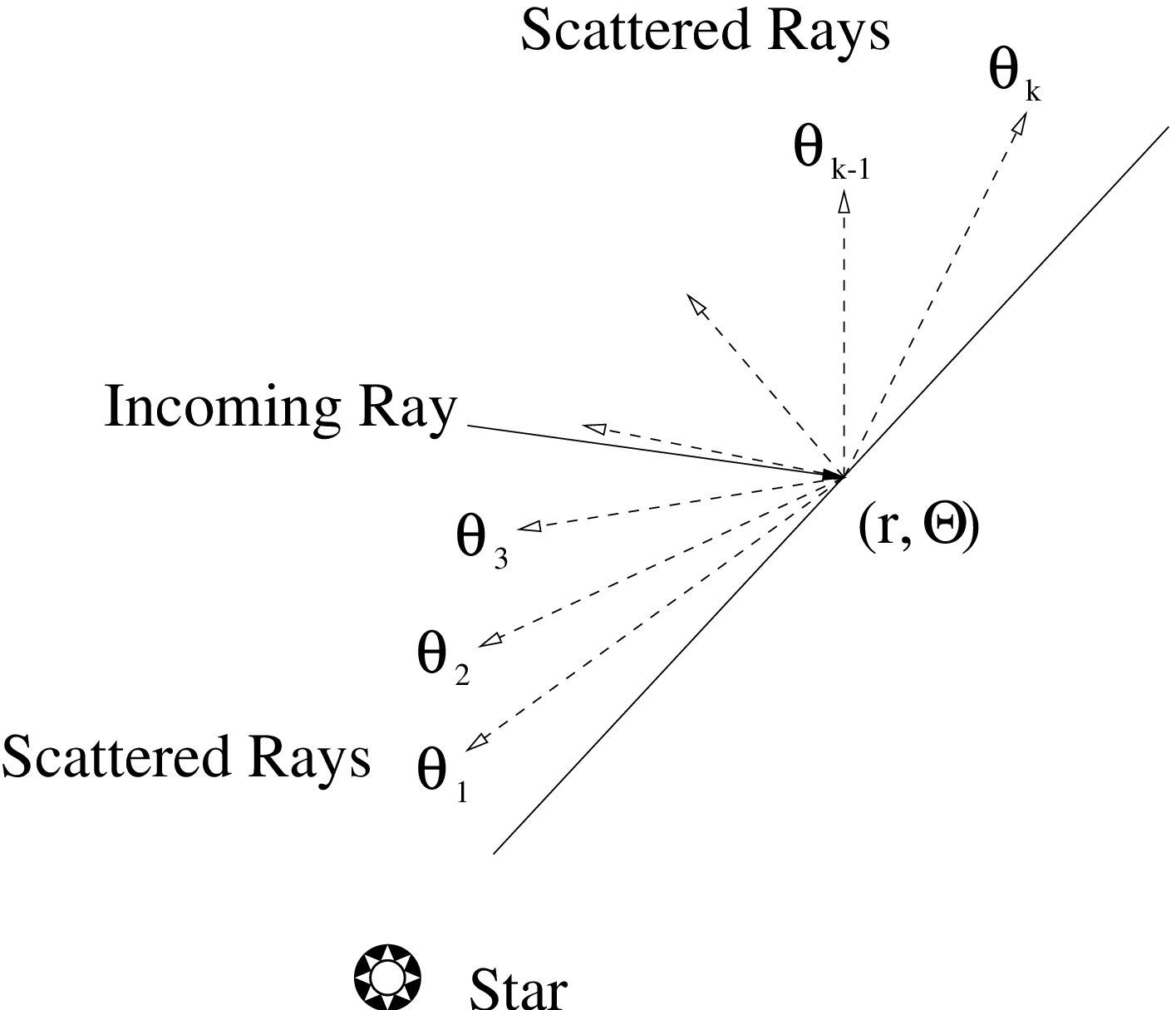}
\hfill
\end{center}
 \figcaption{\label{scatteringimage}%
 [{\sl Left}] 
 The scattering angle, $\omega$, is defined by the acute
 angle between the directions of incident and scattering.
 [{\sl Right}]
 In the anisotropic scattering mode, the incoming 
 angle-integrated intensity available at a given grid is
 redistributed into each of the $\theta$ directions according
 to the weights computed by the scattering phase function.}
\end{figure}

In {\twodust}, anisotropic scattering by dust grains 
is incorporate as follows.
We describe the effect of scattering in the local 
radiation field by generalizing the source function
with the scattering phase function, $\Phi (\omega)$, as
\begin{equation}
 S_{\nu} = \frac{1}{\kappa_{\nu} + \sigma_{\nu}}
  \biggl[ \kappa_{\nu} B_{\nu} 
   \left( T(r, \Theta) \right) + \sigma_{\nu}
   \frac{\int_{4 \pi} I_{\nu} (r, \Theta; \theta, \phi)
   \Phi (\omega) d \omega}{\int_{4 \pi} \Phi (\omega) d \omega}
 \biggr]
\end{equation}
where $\int_{4 \pi} d \omega$ refers to the directional integration 
over $4 \pi$ steradian. 
For the phase function, we employ the modified Henyey-Greenstein 
phase function \citep{cs92} of the form
\begin{equation}
 \Phi (g, \omega) = \frac{3}{2} 
  \left( \frac{1-g^2}{2+g^2} \right)
  \frac{1+\cos^2 \omega}{(1+g^2-2g \cos\omega)^{3/2}}
\end{equation}
in which 
$\omega$ is the scattering angle measured from the angle of incident 
(Figure \ref{scatteringimage}) and
$g$ is the asymmetric parameter that can be computed 
via Mie theory.
We assume azimuthal symmetry of scattering with respect to 
the angle of incident.

When anisotropic scattering is considered, we need to 
redistribute the incoming angle-integrated intensity 
into each of the discrete ray in the $\theta_{k}$ 
direction (Figure \ref{2ddirection}).
Suppose a ray from some direction is incident upon 
the grid ($r_{i}, \Theta_{j}$).
We first calculate a weight for each of the possible 
scattering directions, $\theta_{k}$ with $k$ running 
from $1$ to $n_{\theta}^{(1)}+n_{\theta}^{(2)}+n_{\theta}^{(3)}$, 
based on the scattering phase function and the actual 
scattering angle (symmetry is assumed for the $\phi_{l}$ 
direction).
Then, the incoming angle-integrated intensity is 
redistributed into each scattering direction according 
to the weight (Figure \ref{scatteringimage}).
This process is repeated for all incoming rays until the 
redistributed, scattered intensity for each 
($\theta_{k}, \phi_{l}$) direction is derived.

\section{Shell Properties\label{shell}}

In {\twodust}, the normalized density profile of the shell, 
$\rho(r, \Theta)$, may be defined by users as a Fortran 
function.
Then, the density at the inner radius along the equator, 
$\rho_{\rm min}$, must be determined from the given 
optical depth at the given frequency through
\begin{equation}
 \tau_{\nu} = k_{\nu} \rho_{\rm min} 
  \int_{\rmin}^{\rmax} \rho \left( r, \frac{\pi}{2} \right) dr.
  \label{dens}
\end{equation}
In the above case where there is only one type of dust species, 
the cross section is in units of cm$^2$ and the density is in 
units of cm$^{-3}$.
However, 
the dust shell may have a number of distinct dust layers
to account for, for example, a shell with distinct composition 
layers (e.g., reflecting a possible surface composition change 
of the central star) and a shell having layers with different 
size distributions (e.g., reflecting some dynamical effects such 
as grain-grain collisions).

When multiple composition layers are present, there would be 
a difference in physical characteristics of the shell depending 
on whether the mass density (g cm$^{-3}$) or 
number density (cm$^{-3}$) of the dust grains is assumed to be 
continuous.
For example, the mass density continuity is needed to create a 
single-composition shell having multiple layers of distinct dust 
size distributions (i.e., the number density is expected to be 
discontinuous at the size distribution boundary).
It is reasonable to assume that dust grains aggregate and/or
fragment, for some physical reason, into grains of different 
sizes while keeping the overall mass density unchanged.
On the other hand, the number density continuity is a necessary 
assumption if one sets up a dust shell with multiple composition
layers following the heterogeneous nucleation theory.
Dust grains are thought to form only when there are nucleation 
sites, i.e., the existing grains on which a mantle of differing
composition can form.
If this is the case, the number density remains to be continuous
at the composition boundaries
(i.e., the mass density is expected to be discontinuous).

With multiple composition layers, equation (\ref{dens}) reads
\begin{equation}
 \tau_{\nu} = \rho_{\rm min} \sum_{m=1}^{n} 
  \int_{r_{m-1}}^{r_{m}} k_{\nu}^{(m)} \rho 
  \left( r, \frac{\pi}{2} \right) dr,
\end{equation}
where $r_{m}$ and $k_{\nu}^{(m)}$ are respectively the
outer boundary and the total extinction cross section for
the $m$ th composition layer and $n$ is the total 
number of layers.
Depending on the physical considerations specific to the
problem to be solved, $\rho_{\rm min}$ can be either
the mass or number density.
This distinction, of course, would affect the physical 
meaning of the cross sections.

Once $\rho_{\rm min}$ is determined, the total {\sl dust}
mass of the shell can be calculated by integrating over the entire
shell.
Then, the {\sl rate} of dust mass loss can be determined
by dividing the total dust mass in the shell by the 
age of the shell estimated from the wind crossing time 
in the shell.

\end{document}